\documentclass[a4paper,aps,nofootinbib]{revtex4}
\usepackage{amsmath}
\newcommand{\B}{\mathbf}

\begin{document}

\begin{abstract}
Explicit expressions for the quasi-Kinnersley tetrad for the quasi-Kerr metric are given.  These provide a
very clear and simple example of the quasi-Kinnersley tetrad, and may be useful in the future development of
a `quasi-Teukolsky' scheme for the analysis of perturbation equations in spacetimes which are Petrov type I
but in some sense close to type D.
\end{abstract}

\title{A quasi-Kinnersley tetrad for the quasi-Kerr metric}
\author{Frances White}
\affiliation{Institute of Cosmology and Gravitation, University of Portsmouth, Portsmouth, PO1 2EG, UK}
\date{June 2006}
\maketitle


There has been considerable recent interest \cite{QKI,QKII,QKIII,QKIV,BWMB,LazarusII} in the development and
use of the quasi-Kinnersley tetrad.  This is a Newman-Penrose null tetrad for a Petrov type I spacetime that
will reduce to the Kinnersley tetrad as the spacetime reduces to the Petrov type D Kerr spacetime. There is
any number of quasi-Kinnersley tetrads for a given spacetime and to obtain \textit{a} quasi-Kinnersley tetrad
it is only necessary to find a tetrad that satisfies this criterion \cite{NerozziPrivate}.  Whilst this means
that a quasi-Kinnersley tetrad need not be transverse, i.e. have $\psi_{1}=\psi_{3}=0$ (indeed the
quasi-Kinnersley tetrad given in \cite{QKIV} for the Bondi-Sachs metric is not transverse), transverse
tetrads are generally considered useful (the Weyl scalars $\psi_{1}$ and $\psi_{3}$ are frequently taken to
be non-physical longitudinal effects, whilst $\psi_{0}$ and $\psi_{4}$ will carry the transverse, potentially
radiating, effects), and the quasi-Kinnersley tetrads formally constructed in \cite{QKI} and \cite{QKII} are
all transverse.

In any type I spacetime there are exactly three equivalence classes of transverse Newman-Penrose tetrads,
where equivalent tetrads are related by a class III rotation (a `spin-boost' Lorentz transformation). In only
one of these equivalence classes does the radiation scalar $\xi := \psi_{0}\psi_{4}$ \cite{BeetleBurko}
(defined for a transverse tetrad) go to zero when the speciality index \cite{BakerCampanelli} goes to $1$,
its exact value in a Petrov type D spacetime. Members of this class are defined to be quasi-Kinnersley
tetrads, and hence up to `spin-boost' there is only one \textit{transverse} quasi-Kinnersley tetrad.

The formal construction of the quasi-Kinnersley tetrad being a little involved, and a principal use of the
frame being numerical, a simple explicit example may be helpful to clarify the ideas involved. Existing
explicit examples include the (already mentioned) Bondi-Sachs metric \cite{QKIV}, the Kasner spacetime
\cite{QKIII}, and the Hartle-Thorne metric \cite{FWthesis}.  In addition, \cite{LazarusII} gives and uses a
recipe for constructing the quasi-Kinnersley tetrad for any spacetime obtained using a 3+1 ADM numerical
evolution, assuming a Kerr final state.  Here we give a quasi-Kinnersley tetrad for the recently proposed
\cite{Glamp} quasi-Kerr metric, a particularly appropriate and straightforward case since the original
Kinnersley tetrad is itself for the Kerr metric.

Glampedakis and Babak \cite{Glamp} (henceforth GB) are deliberately agnostic about the source of their
quasi-Kerr metric, it is put forward as a practical scheme which may be useful in the analysis of future LISA
data for assessing the deviation of an observed object from the Kerr metric. In addition, the quasi-Kerr
metric may be useful in the development of a `quasi-Teukolsky' scheme in which the small deviation (of the
quasi-Kerr metric) from the Kerr metric, and from Petrov type D, may allow analysis of the evolution
equations for the perturbed Weyl scalars. (Teukolsky's classic result \cite{Teukolsky} for a type D spacetime
relies on the decoupling and separation of the equations for the perturbations carried by $\psi_0$ and
$\psi_4$.) A quasi-Kinnersley tetrad may be useful in this task.

The quasi-Kerr metric is the Kerr metric with an `extra piece' added to the quadrupole moment, and is
constructed as follows. The Kerr metric is a two parameter (mass and angular momentum) axisymmetric
stationary exact solution of the Einstein Equations, accepted as describing the exterior of a rotating black
hole. The Hartle-Thorne (HT) metric \cite{Hartle1},\cite{Hartle2} is a three parameter (mass, angular
momentum and quadrupole moment) axisymmetric stationary approximate solution of the Einstein Equations for a
rotating (neutron) star, constructed as a truncation at second order of a perturbative expansion in the
rotation rate. Putting one of the three HT parameters equal to a function of another (i.e. relating the
quadrupole $Q$ and angular momentum $J$ by $Q=-J^2/M$) reduces the exterior HT metric to the Kerr metric
truncated at first order in the angular momentum. The \textit{quasi}-Kerr metric can be thought of in two
ways, either that it is the full Kerr metric plus the part of the HT quadrupole term which is additional to
the Kerr quadrupole term, or that it is the HT metric with the all the higher order (in $a$)
Kerr terms added in. GB write \\
\begin{subequations}
\begin{eqnarray}
\label{metricDef}
 g_{ab}^{GB} &=& g_{ab}^{K} +\epsilon h_{ab} +\mathcal{O}(\delta M_{l \geq 4},\delta S_{l\geq 3})\\[0.3cm]
\mbox{where}\hspace{2cm} \epsilon &=& (Q^{K}-Q)/M^3 \\[0.1cm]
Q^{K} &=& -J^2/M
\end{eqnarray}
\end{subequations}
Here $M_{l}$ and $S_{l}$ are the mass and current multipole moments, the $h_{ab}$ are the terms added in
order to modify the quadrupole moment, and superscript $\rm K$ refers to the Kerr metric.  To find the
$h_{ab}$, the HT metric as given in the original papers must first be put into the Boyer-Linquist (BL)
coordinates as used in the Kerr metric; the transformation from BL to HT coordinates is given in
\cite{Hartle2} (note sign error, corrected in \cite{BWMB}) and GB obtain the inverse transformation in a
simple way. The coefficients of the quadrupole moment are then extracted and the parts equal to the
quadrupole terms in the Kerr metric are subtracted. In BL coordinates $\{t,r,\theta,\phi\}$ the result, here
in the covariant as opposed to the contravariant form given by GB, is :
\begin{subequations}
\begin{eqnarray}
h_{tt} &=&  2 (1-2M/r) F_{1}(r) P_{2}(\cos\theta)\\[0.1cm]
h_{\theta\theta} &=& -2 r^2 F_{2}(r) P_{2}(\cos\theta) \\[0.2cm]
h_{rr} &=& (1-2M/r)^{-2}\, h_{tt}  \\[0.1cm]
h_{\phi\phi} &=& \sin^2\theta\, h_{\theta\theta}
\end{eqnarray}
\end{subequations}
where $F_{1}(r)$ and $F_{2}(r)$ are functions of $r$ defined by GB, and $P_{2}(\cos\theta)$ is the usual
Legendre polynomial.  It is important to note that in calculations with the quasi-Kerr metric, all terms of
order $\mathcal{O}( \epsilon a, \epsilon^2)$ are discarded.  Terms such as these have been removed from the
metric definition (equation (\ref{metricDef})) in order to ensure that the quadrupole be the only moment
altered.

To write down a quasi-Kinnersley tetrad for the quasi-Kerr metric we simply add on to all the vectors of the
Kinnersley tetrad for the Kerr metric in BL coordinates a term proportional to $\epsilon$, for example
\begin{subequations}
\begin{eqnarray}
{l_{\rm GB}}^{t} &=& {l_{\rm K}}^{t}+\epsilon p_{0} \\[0.2cm]
{n_{\rm GB}}^{t} &=& {n_{\rm K}}^{t}+\epsilon q_{0} \\[0.2cm]
{m_{\rm GB}}^{t} &=& {m_{\rm K}}^{t}+\epsilon (x_{0}+i y_{0})
\end{eqnarray}
\end{subequations}
where $p_{0}$, $x_{0}$ and $y_{0}$ are functions of $r$ and $\theta$, and subscript $\rm K$ labels vectors of
the Kinnersley tetrad for the Kerr metric. Finding the 16 unknown functions (always discarding terms of order
$\mathcal{O}( \epsilon a, \epsilon^2)$) is straightforward. First, the NP tetrad conditions
\begin{eqnarray}
 \B{l}\cdot \B{n}=-1,\hspace{0.3cm} \B{m}\cdot \bar{\B{m}}=1, \hspace{0.3cm} \mbox{all
other products zero}
\end{eqnarray}
are solved for 10 of the unknown functions. The remaining six free functions correspond to the freedom to
carry out a Lorentz transformation at first order in $\epsilon$. Simply setting all of these six free
functions to zero does result in a quasi-Kinnersley tetrad, but it is not transverse : it does not have
$\psi_{1}=\psi_{3}=0$.  A transverse tetrad can be found by simply calculating the Weyl scalars in the tetrad
that we have so far, with all six free functions non-zero but unspecified, and solving the two complex
equations $\psi_{1}=\psi_{3}=0$ for four of the free functions. The result is that three of the four are set
to zero and the fourth becomes a function of the $h_{ab}$. The remaining two free functions (of $r$ and
$\theta$) are the coefficients of $\epsilon$ in $l^{t}$ and $\Re(m^{\phi})$, and correspond to the freedom to
carry out a class III tetrad rotation (at first order in $\epsilon$) whilst remaining in the equivalence
class of all transverse quasi-Kinnersley tetrads related by class III rotations.

The original Kinnersley tetrad for the Kerr metric is picked out of its equivalence class by setting the spin
coefficient $\epsilon_{\rm sc} = {\scriptstyle\frac{1}{2}} (n^{\mu}l_{\mu;\nu}l^{\nu}+
m^{\mu}\bar{m}_{\mu;\nu}l^{\nu})$ to zero.  Note subscript ${\rm sc}$ to distinguish the spin coefficient
from the perturbation parameter.  Here, calculating $\epsilon_{\rm sc}$ and setting the real and imaginary
parts to zero yields the $r$ dependence of the remaining two free functions (since the equations to solve are
partial differential equations), leaving two free functions of $\theta$ only.  These correspond to the fact
\cite{QKI}, which follows directly from the definition of $\epsilon_{\rm sc}$, that $\epsilon_{\rm sc}$
remains equal to zero under type III rotations with parameters $p$ that both satisfy $l^{a}\nabla p_{a} = 0$.
Here the rotation is at first order, and the condition on the parameters becomes simply $p=p(\theta)$. As yet
lacking a physical interpretation to tie down this remaining freedom, we here put both the remaining free
functions of $\theta$ to zero for simplicity.  The final expressions for the quasi-Kinnersley tetrad for the
quasi-Kerr metric are then
\begin{subequations}
\begin{eqnarray}
\B{l}_{\rm GB} &=& \left[\frac{r^2+a^2}{\Delta}+2\epsilon\,\left(\frac{r}{r-2M}\right)
F_1(r)P_2(\theta),\,\,1,\,\,-2\epsilon\,
\left(\frac{r}{r-2M}\right) q_2(r,\theta),\,\, \frac{a}{\Delta} \right]\\[0.2cm]
\B{n}_{\rm GB} &=& \frac{1}{2}\left[\frac{r^2+a^2}{\Sigma},\, -\frac{\Delta}{\Sigma}+
2\epsilon\,\left(\frac{r-2M}{r}\right)
F_1(r) P_2(\theta),\,2\epsilon\, q_2(r,\theta),\,\frac{a}{\Sigma} \right] \\[0.2cm]
\B{m}_{\rm GB} &=&  \frac{1}{\sqrt{2}}\left[\frac{a\sin\theta(a\cos\theta+ir)}{\Sigma},\,\,2\epsilon\,r
q_2(r,\theta),\,\,  -i\sin\theta\, {m_{\rm GB}}^{\phi},\,\,
\frac{a\cos\theta+ir}{\Sigma\sin\theta}+\epsilon\, \frac{iF_2(r)P_2(\theta)}{r\sin\theta} \right] \\[0.1cm]
\nonumber
\end{eqnarray}
where $\Sigma=r^2+a^2\cos^2\theta$ and $\Delta=r^2-2Mr+a^2$ as usual, and where $q_2(r,\theta)$ is defined to
be
\begin{eqnarray}
q_{2}(r,\theta) &:=& -\frac{5\cos\theta\sin\theta}{32M^3}\left\{3(r-2M)(r+2M)\ln\left(\frac{r}{r-2M}\right)
+\frac{2M}{r^2}\left(6M^3+8M^2r-3Mr^2-3r^3\right)\right\}
\end{eqnarray}
\end{subequations}
The non-zero Weyl scalars in this tetrad are :
\begin{subequations}
\begin{eqnarray}
\psi_{0} &=& 3 \left(1-\frac{2M}{r}\right)^{-1}  \frac{F_{1}(r)\sin^2\theta}{r^2} \,\, \epsilon  \\[0.2cm]
\psi_{2} &=& \frac{M}{(r-ia\cos\theta)^3}+\frac{5}{16} F_{3}(r) P_{2}(\theta) \,\, \epsilon \\[0.2cm]
\psi_{4} &=& \frac{3}{4}\left(1-\frac{2M}{r}\right) \frac{F_{1}(r)\sin^2\theta}{r^2} \,\, \epsilon
\end{eqnarray}
where
\begin{eqnarray}
F_{3}(r) &:=&
-\frac{1}{M^2}\left\{3\ln\left(\frac{r}{r-2M}\right)-\frac{2M}{r^4}\left(3r^3+3Mr^2+4rM^2+6M^3\right)\right\}
\end{eqnarray}
\end{subequations}
It is immediately obvious that as the spacetime approaches the type D Kerr spacetime (i.e. $\epsilon
\rightarrow 0$), the vectors of the quasi-Kinnersley tetrad approach those of the Kinnersley tetrad and the
Weyl scalars approach the known expressions for the scalars in the Kinnersley tetrad of the Kerr metric, i.e.
the only non-zero Weyl scalar is $\psi_{2} = M/(r-ia\cos\theta)^3 $.  It is straightforward to check that all
the spin coefficients also approach their Kerr expressions.

The speciality index $S$ \cite{BakerCampanelli} is a tetrad invariant quantity equal to $1$ (only) in a
Petrov type D spacetime.  In a transverse tetrad it can be calculated as
\begin{eqnarray}
S = 27\psi_{2}^2 \left(\psi_{0}\psi_{4}-\psi_{2}^2\right)^{2}\left(\psi_{0}\psi_{4}+3\psi_{2}^2\right)^{-3}
\end{eqnarray}
Here we get
\begin{eqnarray}
S = 1-\frac{27}{4} \frac{r^2 F_1(r)^2\sin^4\theta}{M^2}\,\epsilon^2 +\mathcal{O}(a\epsilon^2,\epsilon^3)
\end{eqnarray}
This expression for $S$ for the quasi-Kerr metric is in fact the same as that for the Hartle-Thorne metric
given in \cite{BWMB}; recall that $\epsilon = (Q^{\rm Kerr}-Q)/M^3$ and note that the coordinate
transformation from BL to HT coordinates is not needed here since, as GB point out, either set of coordinates
can be used in $\mathcal{O}(J^2,Q)$ terms. In fact, moving to the symmetric tetrad (using a class III
rotation to put $\psi_{0}=\psi_{4}$), truncating terms zeroth order in $\epsilon$ to first order in $a$, and
using the coordinate transformation as appropriate, we recover the expressions for the Weyl scalars and
vectors of the quasi-Kinnersley tetrad for the Hartle-Thorne metric as given in \cite{BWMB} and
\cite{FWthesis} respectively \footnote{To carry out these checks it is necessary to note that \cite{BWMB} and
\cite{FWthesis} use $Q=-M_{2}=J^2/M$ while GB use $Q=M_{2}=-J^2/M$, where $M_2$ is the $l=2$ mass multipole
moment, and also that while the signature of the metric in both papers is given as $-+++$, the Weyl scalars
given in \cite{BWMB} and \cite{FWthesis} have the signs that would arise using the opposite signature and
those in the current paper do not.}.  We have hence found a far more straightforward and transparent route to
the explicit expressions for these quantities than previously used.

Note that in the expression for $S$ the first non-zero term in $\epsilon$ has been retained despite the fact
that it is higher order than terms retained in the metric components and all other quantities we have
calculated (it is straightforward to check that the addition of terms in $a\epsilon$ and $\epsilon^2$ to the
Weyl scalars does not alter the result for $S$).  This form of $S$ arises simply because the definition of
$S$ is quadratic in the Weyl scalars and because here $\psi_{0}$ and $\psi_{4}$ have no zeroth order terms.
As pointed out in \cite{BWMB}, an $S$ of $1$ at first order does not necessarily imply that the spacetime is
type D at first order, since there may be four distinct principal null directions at this order. (An example
of this is given in \cite{FWthesis} where an explicit calculation shows that at first order there are indeed
four distinct principal null directions for the Hartle-Thorne metric, while the speciality index deviates
from $1$ only at second order. See also \cite{Cherubini}.)


 \end{document}